# The future of statistical disclosure control

**December 2018**

Professor Mark Elliot
University of Manchester

Professor Josep Domingo-Ferrer
Universitat Rovira i Virgili

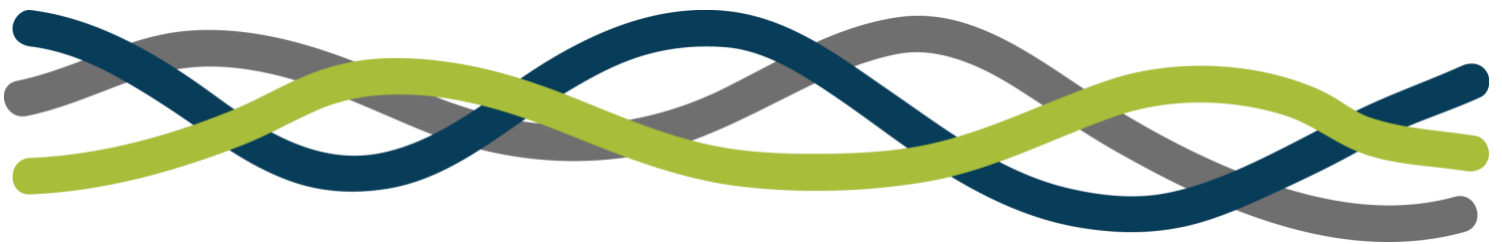

**About National Statistician's Quality Reviews (NSQR)**

National Statistician's Quality Reviews (NSQRs) cover thematic topics of national importance, conducted on behalf of and for the Government Statistical Service (GSS) in the United Kingdom. These reviews are future facing, ensuring that methods used by the GSS are keeping pace with changing data sources and technologies.

Articles contributed by leading experts published as part of the Privacy and Data Confidentiality NSQR contribute to the continuous improvement of these methods and support the GSS identify what good practice looks like for these methodologies as well as help identify opportunities for further development and investment. More information can be found on the Government Statistical Service website at: https://gss.civilservice.gov.uk/guidances/quality/nsqr/privacy-and-data-confidentiality-methods-a-national-statisticians-quality-review/. For further enquiries please contact Gentiana D. Roarson at gentiana.roarson@ons.gov.uk or the GSS Quality Centre at qualitycentre@statistics.gov.uk.

**Authors**

Professor Mark Elliot, School of Social Sciences, University of Manchester, Oxford Road Manchester M13 9PL, UK; Email: mark.elliot@manchester.ac.uk

Professor Josep Domingo-Ferrer, Department of Computer Engineering and Mathematics, Universitat Rovira i Virgili, Av. Països Catalans 26, E-43007 Tarragona, Catalonia. E-mail: josep.domingo@urv.cat

**Contents**

## 1. Introduction[1]

Statistical disclosure control (SDC) was not created in a single seminal paper nor following the invention of a new mathematical technique, rather it developed slowly in response to the practical challenges faced by data practitioners based at national statistical institutes (NSIs). SDC's subsequent emergence as a specialised academic field was an outcome of three interrelated socio-technical changes: (i) the advent of accessible computing as a research tool in the 1980s meant that it became possible - and then increasingly easy - for researchers to process larger quantities of data automatically; this naturally increased demand for such data; (ii) it became possible for data holders to process and disseminate detailed data as digital files and (iii) the number of organisations holding data about individuals proliferated. This also meant the number of potential adversaries with the resources to attack any given dataset increased exponentially.

In the remainder of this article, we will describe the state of the art for SDC and then discuss the core issues and future challenges. First though, we want to clarify the distinction between SDC and two other terms which are often incorrectly interchanged with it: *anonymisation* and *privacy*.

### 1.1  SDC and anonymisation

Anonymisation is a process whereby personal data is transformed into non-personal data (also then often referred to as "anonymised data"). Anonymisation is especially attractive because provisions and restrictions on the processing of personal data by the EU's General Data Protection Regulation (and other laws which regulate of data processing) do not apply to anonymous data.

SDC methods are used as part of anonymisation processes. They attempt to control/limit the risk of re-identification and attribute disclosure (see Section 2.1.1) through manipulations of the data. However, anonymisation processes involve more than just SDC. Firstly, approaches such as privacy models can be considered as alternative anonymisation processes – and, secondly, any fully functional anonymisation process must also consider things such as context, provenance, external communications and impact (because risk is not just about probability).

---

[1] Acknowledgement: Some of the material in this document is drawn from other work by the authors (open access publications, research proposals etc.). The authors thank the co-authors of those sources Jennifer Taub, Joe Sakshaug, Elaine Mackey, Duncan Smith and Dave Martin) for consenting to this re-use.

Therefore SDC is just one of the tools in the anonymiser's toolbox; it is neither necessary nor sufficient for functional anonymisation but it is extremely useful [1].

## 1.2 Disclosure control is not privacy

The word privacy is a much abused and hugely contested term, and an important distinction should be made between confidentiality (which concerns data) and privacy (which concerns people) [2]. This is not the appropriate place for philosophical discussion [3], but it is important to realise that SDC (and indeed all processes which attempt to manipulate disclosure risks) are primarily technical confidentiality processes which only indirectly and imperfectly map onto privacy protections.

The key point is that one should not assume that one has adequately protected privacy by controlling disclosure. Even the so called "privacy models" (which we will discuss later) do not achieve this. To fully account for privacy concerns there is a need to ethically audit the data processing much more deeply and thoroughly than what is required for even well-formed disclosure control. It is true that disclosure control may have a positive impact on privacy, but that impact is a secondary by-product of its primary role as a confidentiality process. SDC is, and can only be, a mechanism for meeting confidentiality assurances.

We will use the term privacy sparingly in this article. We will mainly do so where we refer to the work of others who have used the term to avoid confusing the reader. For example, we will not rename "differential privacy" and will use the term "privacy guarantee" despite believing these terms to be misnomers. However, the reader should note that the term "privacy" is in effect standing in for "confidentiality" and that real privacy is not in fact being discussed.

## 2. The state of the art

## 2.1 Basic principles

SDC fundamentally consists of two processes: disclosure risk analysis and disclosure control. Controlling the disclosure risk must be done in a way that optimises the trade-off between risk and utility. While risk must be kept below the maximum acceptable threshold (set by law or by good practices), utility must be kept above the minimum threshold that data users can accept. Without utility constraints, there would be no reason to control disclosure: one might rather suppress the data entirely, which would result in 0% disclosure risk!

### 2.1.1 Attribution and identification

There are two main risks of disclosure that need to be controlled:

- *Attribute disclosure* occurs when the intruder is able to estimate the value of a confidential attribute for a certain subject. Note that some parts of the literature distinguish between attribute and inferential disclosure (with the former absolute, and the latter probabilistic) but we consider this to be a false distinction as all such disclosures are moderated by error processes.
- *Re-identification disclosure* occurs when an intruder can determine the identity of the subject to whom a record or a data item corresponds.

### 2.1.2 Input vs output SDC

Another key distinction is between input and output disclosure control. Input disclosure control is control placed on the data before it is analysed (or even seen by an analyst). Output SDC places controls on analytical outputs. Formally, output SDC is most commonly used in situations such as research data laboratories where it is possible to control risk through environmental controls and thereby allow access. We shall have more to say about output disclosure control later as there are some issues with it that need attention; for the remainder of this section however we will focus on input SDC.

### 2.2 Controlling disclosure risk: ex-post vs privacy models

The traditional approach to anonymisation, still dominant among statistical agencies, controls disclosure risk ex-post: an SDC method with a heuristic parameter choice and with suitable utility preservation properties is run on the data and, *after that*, the risk of disclosure is measured[2].

For example, the risk of re-identification can be estimated empirically by attempting record linkage between the original and the anonymised datasets [4], or analytically, by using generic measures [5, 6] or measures tailored to a specific anonymisation method [7]. If the remaining risk is deemed too high, the anonymisation method must be re-run with more stringent parameters or a further method applied (probably with more utility sacrifice).

---

[2] Procedurally, it is an iteration between risk assessment and control. Potentially, a risk analysis is the first step to determining whether controls are necessary and the extent to which they are required, but risk assessment must necessarily be the last step. The key point is that there is a separation of the steps of quantifying the risk and controlling it.

The alternative anonymisation approach, introduced by computer scientists, controls risk ex-ante and is based on the notion of *privacy model*, which is a condition, dependent on a parameter, that guarantees an upper bound on the risk of re-identification disclosure and perhaps also on the risk of attribute disclosure by an intruder. The model can be enforced using one or several SDC methods whose parameters derive from the model parameters.

### 2.2.1 Privacy models

The first privacy model to be developed was *k*-anonymity [8]. It focuses on quasi-identifier (QI) attributes, that is, on attributes that are not direct identifiers (recall that anonymised datasets cannot contain direct identifiers), but that in combination might be used to link the anonymised data with an external identified data source to re-identify the subject to whom an anonymised record corresponds. Examples of quasi-identifiers include *Age, Profession, Gender* and *Zipcode*. Under *k*-anonymity, QI attribute values are generalised or microaggregated (see Section 2.4.4), so that each combination of QI attribute values is shared by at least *k* records (the so-called *k*-anonymous class); in this way, the risk of re-identification is at most *1/k*. K-anonymity can either be implemented using a combination of generalisation and local suppression [8], or be based on microaggregation [9]. Unfortunately, *k*-anonymity fails to guard against attribute disclosure. Indeed, it can occur that the values of a confidential attribute (e.g. *Salary*) are very similar for all the records in a *k*-anonymous class; in that case, even if the intruder cannot determine the record corresponding to his target subject, he can obtain a good guess of the target subject's salary.

Extensions of *k*-anonymity that protect against attribute disclosure include *l*-diversity [10], *t*-closeness [11], *(n,t)*-closeness [12], crowd-blending privacy [13], $\beta$-likeness [14] and others.

The introduction of $\varepsilon$-differential privacy (DP) inaugurated a family of privacy models essentially different from *k*-anonymity, with the attractive feature that they make no assumptions about the intruder's auxiliary knowledge (e.g. external identified data sources, extra knowledge about target subjects, etc.) [15]. The anonymised response to a query is $\varepsilon$-DP if it does not betray (up to a function of $\varepsilon$) the presence or absence of any subject in the original dataset. Although DP was designed to anonymise query results, methods to produce DP datasets can also be found in the literature with proposals for applications to histograms [16], microaggregation [17], Bayesian networks [18] and synthetic data [19, 20].

*K*-anonymity-like models and $\varepsilon$-DP should not be regarded as competing models: each has its strength and weakness [21, 22, 23] and indeed each can be used to boost the other [24]. Although DP is a prevalent approach among the academic community, research on other privacy models is also needed given the diversity of requirements of controllers and users.

## 2.3 Intruder testing

Intruder testing (sometimes called penetration testing) is an approach to assessing disclosure risk more directly than the range of methods described above. Essentially, the tester attempts to mimic a 'would-be' data intruder by re-identifying units within a dataset. That simple description hides a complex and resource-intensive process. Here we just provide a brief description to intruder testing (see [1] for a more comprehensive description):

There are essentially four stages to an intruder test: (i) data gathering; (ii) data preparation and harmonisation; (iii) the attack itself; and (iv) verification. The first stage tends to be the most resource-intensive and the second and third require the most expertise.

Data gathering involves going out to the world and gathering information on particular individuals. Exactly what that will look like will depend on the nature of the scenario tested but would typically involve at least some searching of the Internet. For example, one intruder test in the literature gathered information on 100 individuals, taking about three person-months of effort [25]. This test also included a second augmented attack using data purchased from the commercial data broker CACI [25].

Once the data gathering phase is complete, the data have to be harmonised with the target dataset. This will require work both across all the data, and at the level of individual records, as in all likelihood there will be several issues to be addressed to achieve this. For example, gathered data will often be coded differently from the target data; the tester might have gathered information about somebody's job from social media, but how exactly would that be coded in the target dataset? There will also be *data divergence* within the gathered information.

Some scenarios simulate linkage between an identification dataset and a target dataset, rather than between gathered data and a target dataset. In this case no data gathering is needed but data harmonisation will still usually be required and issues of data divergence still be critical, although the focus will tend to be on the dataset as a whole rather than upon individual records.

The details of the attack stage will also depend on the nature of the data and the intrusion scenario, but typically it will involve attempting to link the information gathered at stage (i) to the tester's dataset. Usually this will involve a mixture of automated and manual processes. In essence the aim is to establish negative and positive evidence for links between the attack information and records in the dataset.

A second point to note is that no match has 100% confidence associated with it. This reflects the reality that we can never be completely certain that we are correct. There is always a possibility that (i) the dataset contains data for a person who is highly similar to our target – their statistical twin – or that (ii) the assumption that our target is in the data is incorrect. It is worth noting in passing that this is the flipside of not being able to reduce the risk to zero.

Finally, once the matches have been selected, they need to be verified. This will often be carried out by a different person or organisation than the person doing the matching. If the matcher is carrying it out – at the risk of stating the obvious – they should only do this once they have decided upon their final list of matches.

As mentioned above intruder testing done properly is resource-intensive – it can be very informative but is recommended only for new data situations where the calibration of statistical disclosure risk measures is difficult to achieve [1].

## 2.4 SDC methods

SDC methods can be classified depending on the data format to which they are applied. We will distinguish three main data formats: tables, database query outputs and microdata.

### 2.4.1 SDC methods for tables

Methods to control disclosure in tables can be classified as:

- *Non-perturbative.* They do not modify the values in the cells, but they may suppress or recode them. Best known methods: cell suppression (CS) and recoding of categorical attributes. Cell suppression has a long tradition in national statistical offices: a sensitivity rule (for example, the dominance rule or the $p$% rule) is used to identify cells that are sensitive; these sensitive cells are suppressed (primary suppressions) and then another set of cells (called secondary suppressions) is identified and suppressed to prevent inferring the primary suppressions from table marginals.
- *Perturbative.* They modify the values in the cells. Best known methods: controlled rounding (CR) and controlled tabular adjustment (CTA). These methods also rely on identifying sensitive cells, but they have the advantage of yielding a protected table without missing cells. Their shortcoming is that the reported cell values may not correspond to the true cell values.

More details on those tabular SDC methods, and the τ-Argus [27] package for implementations of them, can be found in the literature [26].

**2.4.2 SDC methods for database query outputs**

There are two main SDC principles for queryable database protection:

- *Query perturbation.* Perturbation (noise addition) can be applied to the microdata records on which queries are computed (input perturbation) or to the query result after computing it on the original data (output perturbation). Differential privacy (see Section 2.1.1) was initially proposed as an output perturbation method for query outputs.
- *Query restriction.* The database refuses to answer certain queries. A common criterion to decide whether a query can be answered is the query set size: the answer to a query is refused if this query together with the previously answered ones isolates too small a set of records. The main problems of query restriction are: i) the computational burden to keep track of previous queries; ii) collusion attacks can circumvent the query limit. The issue of trackers - a sequence of queries to an on-line statistical database whose answers disclose the attribute values for a small subgroup of individual target records or even a single record - is well known [28]. It has been demonstrated that building a tracker is feasible and not resource-intensive for any subgroup of target records [29].

**2.4.3 SDC methods for microdata**

A microdata file *X* with *s* respondents and *t* attributes is an *s* × *t* matrix where $X_{ij}$ is the value of attribute *j* for respondent *i*. Attributes can be numerical (e.g. *Age, Blood Pressure*) or categorical (e.g. *Gender, Job*). Depending on their disclosure potential, attributes can be classified as:

- *Identifiers.* Attributes that unambiguously identify the respondent (e.g. *Passport Number, Social Security Number, Name-Surname,* etc.).
- *Quasi-identifiers or key attributes.* They identify the respondent with some ambiguity, but their combination may lead to unambiguous identification (e.g. *Address, Gender, Age, Landline Telephone Number*, etc.).
- *Confidential outcome attributes.* They contain sensitive respondent information (e.g. *Salary, Religion, Diagnosis*, etc.).
- *Non-confidential outcome attributes.* Other attributes which contain non-sensitive respondent info.

Identifiers are of course suppressed in anonymised datasets. Disclosure risk comes from quasi-identifiers (QIs), but these cannot be suppressed because they often have high analytical value. Indeed, QIs can be used to link anonymised records to external non-anonymous databases (with identifiers) that contain the same or similar QIs; this leads to re-identification. Hence, anonymisation procedures must deal with QIs.

There are two principles used in microdata protection, data masking and data synthesis:

- Masking generates a modified version *X′* of the original microdata set *X*, and it can be *perturbative masking* (*X′* is a perturbed version of the original microdata set *X*) or *non-perturbative masking* (*X′* is obtained from *X* by partial suppressions or reduction of detail, yet the data in *X′* are still true).
- Synthesis is about generating synthetic (i.e. artificial) data *X′* that preserve some pre-selected properties of the original data *X*.

### 2.4.4 Perturbative masking

The main principles for this class of masking are:

- *Noise addition.* This principle is only applicable to numerical microdata. The most popular method consists of adding to each record in the dataset a noise vector drawn from a *N(0,αΣ)*, with *Σ* being the covariance matrix of the original data. Means and correlations of original data can be preserved in the masked data by choosing the appropriate *α*. Additional linear transformations of the masked data can be made to ensure that the sample covariance matrix of the masked attributes is an unbiased estimator for *Σ*.
- *Microaggregation.* Microaggregation partitions records in a dataset into groups containing each at least *k* records; then the average record of each group is published [30]. Groups are formed by the criterion of maximum within-group similarity: the more similar the records in a group, the less information loss is incurred when replacing them by the average record. There exist microaggregation methods for numerical and also categorical microdata [9, 31].
- *Data swapping.* Values of attributes are exchanged among individual records, so that low-order frequency counts or marginals are maintained. Although swapping was proposed for categorical attributes, its rank swapping variant is also applicable to numerical attributes. In the latter, values of each attribute are ranked in ascending order and each value is swapped with another ranked value randomly chosen within a restricted range (e.g. the ranks of two swapped values cannot differ by more than *p*% of the total number of records).
- *Post-randomisation.* The PRAM method works on categorical attributes [32]: each value of a confidential attribute is stochastically changed to a different value according to a prescribed Markov matrix.

### 2.4.5 Non-perturbative masking

Principles used in this kind of masking include:

- *Sampling.* Instead of publishing the original data file, only a sample of it is published. A low sampling fraction may suffice to anonymise categorical data (the probability that a sample unique is also a population unique is low). For continuous data, sampling alone does not suffice.
- *Generalisation.* This principle is also known as coarsening or global recoding. For a categorical attribute, several categories are combined to form new (less specific) categories; for a numerical attribute, numerical values are replaced by intervals (discretisation).
- *Top/bottom coding.* Values above/ below, a certain threshold are lumped into a single top/ bottom, category respectively.
- *Local suppression.* Certain values of individual attributes are suppressed in order to increase the set of records agreeing on a combination of quasi-identifier attributes. This principle can be combined with generalisation.

## 2.5    Data synthesis

Data synthesis is sometimes viewed as an SDC method and sometimes as an alternative to SDC. This is mostly a semantic concern, but it is true that synthesis has some functional properties different from those of other SDC methods. Yet, the goal is the same: to release a useful dataset whilst maintaining data subject confidentiality.

The concept of synthetic data was first introduced by Donald Rubin, who proposed multiply imputing a whole dataset, so that no real microdata would be released [33]. Soon after, Rob Little proposed an alternative where only the sensitive variables were synthesised, referred to as partially synthetic data [34]. Since fully synthetic data does not contain any original data, the disclosure of sensitive information from the synthetic data is much less likely to occur. Likewise, for partially synthetic data, the sensitive values are synthetic, and thus disclosure of sensitive information is also less likely to occur compared to the original data. Nonetheless, in the (in principle unlikely) event that some or all of the synthetic records are very similar to certain original records, re-identification or attribute disclosure may *de facto* occur. In this situation, there is little chance that the affected subjects will be satisfied by the argument that the data are synthetic and therefore do not denote them. Therefore, it remains good practice to assess disclosure risk for synthetic data before release as one would for masked data [35].

Special care must be devoted to make sure synthetic data yield valid statistical analyses. Validity is important since synthetic data can be used to study policy-relevant outcomes and inform policy decisions; for example, the US Census Bureau released a synthetic version of the Longitudinal Business Database [36, 37] and in Germany a synthetic version of the IAB Establishment Panel has been released [38, 39]. Both of these datasets provide relevant information on businesses in their respective countries, and the quality of these data are of great importance to data users. If the synthetic data produces results that are distorted, incorrect economic conclusions could be drawn. It has been argued that the validity of synthetic data is dependent on the models used to generate them and will not reflect relationships that are present in the original data but not represented in the data generation model [40]. Furthermore, if the distributional assumptions built into the model are incorrect, then these incorrect assumptions will also be built into the user's analysis models. The importance of this issue is exacerbated by the fact that the vast majority of utility tests for synthetic data are not necessarily representative of the work that data analysts intend to undertake. The other side of the coin is that, if too accurate a model is used for synthesis (what is called model overfit), then the resulting synthetic values may be too similar to certain original values, which may result in disclosure (see previous paragraph).

The initial proposal for producing synthetic data was based on multiple imputation (MI) techniques using parametric modelling. Originally, MI was created in the 1970s as a solution to deal with missing data by replacing missing values with multiple values, to account for the uncertainty of the imputed values.

Recent research has examined non-parametric methods - including machine learning techniques - which are better at capturing non-linear relationships for generating synthetic data [41]. These methods include classification and regression tress (CART) [42], random forests [43], bagging [41], support vector machines [44], and genetic algorithms [45]. CART, originally developed as a non-parametric modelling tool based on decision trees [46, 47], has become the most commonly used non-parametric method for generating synthetic data. For example, CART was used to generate synthetic data for some of the variables in the US Longitudinal Business Database [48].

To generate synthetic data from a CART model, each variable is fitted to a tree that splits into branches based on a series of binary splits [42]. These branches continue to divide until they terminate in leaves. The values in the terminal leaf represent the distribution of the predicted variable. The synthetic data is then generated by sampling from the leaves. Previous work found that when comparing different methods of machine learning for synthetic data generation, CART yielded the highest data utility [49]. They tested the utility by comparing distributions of different variables and coefficient estimates from a fitted logistic regression model. Likewise, a utility comparison of different tree-based synthetic datasets against a parametric synthetic dataset found that CART performed better than other tree-based synthetic data approaches, but not as well as the parametric method [50]. This was determined by comparing coefficient estimates from a linear regression, a logistic regression, and Poisson regression. Previous studies tend to be limited in their assessment of data utility in that they typically choose only a limited number of analytical models and outcomes, and additionally they tend to analyse said models using a single metric (e.g. confidence intervals).

CART synthetic data was originally designed under a multiple imputation framework, and the utility of this framework has been evaluated in other work [50, 51]. However, given that CART creates synthetic data using a different ethos from parametrically multiply imputed data, it has many advantages including that CART models (i) are more easily applied especially to non-smooth continuous data, and (ii) provide a semi-automatic way to fit the most important relationships in the data.

## 2.6 Measuring utility and its loss

As justified in the introduction to this article, measuring the utility of SDC-protected data is an integral part of disclosure control. Utility measurement is difficult because in many cases there is no clarity about what the users of data will want to do with those data. There is a distinction between *information loss,* which attempts to capture in information-theoretic terms of the change in information caused by disclosure control, and *utility measurement,* which attempts to get as close as possible to the actual use case for the data. The distinction between the two is not clear-cut and here we refer to *utility loss* as an overarching term for all such methods.

### 2.6.1 Utility evaluation in tabular SDC

For cell suppression, utility loss can be measured as the number or the pooled magnitude of secondary suppressions. For CTA or CR, it can be measured as the sum of distances between true and perturbed cells. The above loss measures can be weighted by cell costs if not all cells have the same importance[3].

### 2.6.2 Utility evaluation in SDC of database queries

For query perturbation, the difference between the true query response and the perturbed query response is a measure of utility loss; this can be characterised in terms of the mean and variance of the added noise (ideally the mean should be zero and the variance small). For query restriction, utility loss can be measured as the number of refused queries.

---

[3] As to disclosure risk, it is normally evaluated by computing feasibility intervals for the sensitive cells (via linear programming constrained by the marginals). The table is said to be safe if the feasibility interval for any sensitive cell contains the protection interval previously defined for that cell by the data protector.

### 2.6.3 Utility evaluation in microdata SDC

Utility loss in disclosure-controlled microdata can be evaluated using either data use-specific loss measures or generic loss measures. The former evaluates the extent to which SDC affects the output of particular analyses [52]. Often, the data protector has no idea regarding what the users will do with the data; if this is so then generic utility loss measures - that measure the impact of SDC on a collection of basic statistics (means, covariances, correlations etc. [53]) or that rely on some score (such as propensity scores [54]) - can be used. The most general methods measure the distance between the original and disclosure-controlled data using measures such as Jensen Shannon divergence [55]. These, however have issues when dealing with data sparsity.

In general, our understanding of data utility is less well developed than our understanding of disclosure risk. One particular issue is that all of the above measures of utility are relative measures; that is, they (attempt to) measure the utility of the data *compared with the original data.* This sets the original data up as an operational gold standard – which is at best a proxy measure and may be completely unrealistic. A 50% reduction in utility may be completely fine if the original data utility was very high and may render the data effectively useless if the original data utility was low. Relatedly, the original data will usually be subject to multiple error processes which clearly impact on their underlying utility. Therefore, in short, we need to better understand what data utility is in and of itself; arriving at such a general theory of data utility would be extremely beneficial for the field.

## 3. Issues arising

Most of the issues that arise in what might be called the 'standard SDC model' are due to the rapid and ongoing change in the global socio-technical system. Specific reasons are the exponential increase in available data as well as a less rapid but nevertheless consistent increase in our capacity and resources to process that data in new, imaginative and valuable ways.

### 3.1  Understanding the data environment

The term "data environment" semi-formalises the notion of the context in which data sits [56]. Underlying all of this is the relationship between the data in this environment and the data that is the focus of a potential release or share. Many authors have commented that this environment is inherently difficult - if not impossible - to understand and therefore directly assessing risk is itself impossible. This in turn has led to bad decision making about data sharing (a strange mixture of over-caution and imprudence which is driven more often than not by the personality of the decision maker rather than by rational processes).

### 3.1.1 Data Environment Analysis

The presupposition of the Data Environment Analysis is that data that are held in private databases are a key source of uncertainty in the SDC process [56]. However, the instruments (e.g. loyalty card applications, service registrations, social media platform sign ups) that are used to collect these data are usually available on the web in electronic form. Data Environment Analysis was developed at the University of Manchester and works by collecting these instruments manually and adding them to a metadata repository. This repository is in turn used to generate the empirically grounded key variable patterns that are needed to perform risk assessments. Although the repository does not contain records, it does enable the possibilities for linkage to be identified, because it creates a *synthetic data environment* in an accessible form, thereby allowing the simulation of intruders who combine multiple (public and private) data sources together to produce attack vectors.

## 3.2  The use of multiple data sources

The increasing diversity of datasets within the global data environment has thrown up another opportunity for analysts to perform the linkage or matching of data to allow more comprehensive analyses. There are numerous processes which fall under this heading, but the paradigmatic ones are *record linkage* and *statistical matching*.

The goal of record linkage is to identify pairs of records *(a, b)* from databases *A* and *B* that relate to the same population units. The classical method for linking records is based on a Bayesian latent model and uses comparisons for equality on the key variables [57] (variables common to *A* and *B*). The basic output of a linkage exercise is a set of link probabilities, one for each *(a, b)* pair.

The goal of statistical matching is to generate a database of records over the union of the variables in *A* and *B* –with no general expectation that merged pairs of records correspond to identical population units [58]. The basic output is a list of plausible full records and weights. The full records are constructed by merging record pairs that are 'similar' on the key variables. The results of record linkage or statistical matching might be used for subsequent analysis.

Recent developments have, to some degree, blurred the distinction between record linkage and statistical matching. Using similarities on the key variables can significantly improve linkage performance [59]. Simultaneously estimating the joint distribution over all the variables (essentially, the goal of statistical matching) has also produced improvements in record linkage performance [60].

The use of linked administrative data for empirical, and particularly policy-oriented, research has increased steadily in recent decades [61, 62]. There are at least two factors driving this trend: 1) a desire to use more detailed and up-to-date data to inform policy and practice; and 2) a gradual change in the political and data infrastructure landscape to facilitate access to record-level data. While linking and expanding access to these data undoubtedly increases scientific research opportunities and facilitates the study of complex policy-oriented research questions, both factors present methodological and data confidentiality issues that have not been fully worked out.

From a confidentiality perspective, the re-use of administrative records for research purposes poses strong risks because they tend to capture whole populations, often contain sensitive data, and consent of the data subjects has not and - pragmatically - could not be obtained. These risks increase further when the data are linked. For these reasons, access to such data is normally restricted to Approved Researchers working within secure laboratory environments with linkage carried out by trusted third parties (TTPs - organisations that facilitate linkage between data owned by different organisations that cannot share data with one another). The secure laboratory model described above was originally designed for single datasets and the output disclosure methods have been designed with that in mind.

### 3.3 The informal nature of output SDC

There has been a lot of research on input SDC covering risk assessment, risk control methods and the measurement of information loss. However, output SDC is under-researched. The methods employed are based on a small number of arbitrary heuristic rules such as the threshold rule. These tend to be informally justified only through a "test of time" assertion. Formal evaluation and testing of these rules seem overdue.

A specific and unexplored issue for output disclosure concerns the situation where the input data are a set of linked records. Where data from multiple owners is combined using a TTP or similar mechanism there is now a possibility that one of the data owners may be able to re-identify units within any given analytical output. Owners will have strong knowledge about which data units will be represented in any given output and knowledge about the specific values for those units in respect of the data that they have contributed. This represents quite a strong adversary; at present no work has been done to assess this particular risk. In general, risk assessment for output SDC needs to be put on a more formal footing.

## 4. The future of SDC

### 4.1 Measuring attribute/inferential disclosure

As mentioned in Section 2, statistical disclosure arises out of attribution and identification. Despite this, SDC as a field has focused heavily on the re-identification risk, an oddity that has given little attention to the attribute disclosure risk.

Two methods that have been developed are the SAP method for tables of counts [63] and the CAP method for microdata [64]. The premise of SAP is that formally a table of counts is disclosive if an intruder who has information about some (>0) population units included in the table can potentially create a zero by removing the known units. The SAP metric is parametrised by the amount of information that the intruder is assumed to have and the output is a probability that an intruder with that amount of information will recover a zero. The CAP method on the other hand assumes that the intruder knows some information about an individual who is (or indeed is not) contained within some microdata.

More work needs to be done to explore attribute disclosure.

### 4.2 SDC and big data

Thus far we have avoided the use of the term *big data*. This is partly because it is a buzzword with no agreed meaning and partly because it is a bit of a semantic bucket into which a people happily toss all manner of socio-technical changes which might be more aptly described and better understood on their own. From a SDC point of view much of what is meant by "big data" can be captured by the opening paragraphs of this article: a set of socio-technical changes that transform the scale, use, meaning and value of data. However, the key point is that we have reached a phase in this process where we cannot deal with this by doing more of the same. Challenges to protect big data against disclosure and to protect data (big or small) in the context of big data include:

- *Unjustified de facto trust in controllers*. Twenty years ago, national statistical institutes (NSIs) and a few others were the only data controllers that were explicitly gathering data on the citizens, and their legal status made them trusted. In contrast, in the current big data scenario, there is a host of controllers gathering and processing information, and it is no longer reasonable to take it for granted that the subject trusts all of them to keep her data confidential and/or anonymise them properly in case of release [65]. Thus, to preserve her confidentiality and, more generally, her informational self-determination, *the subject has to be empowered by giving her agency over her own data*. Local anonymisation gives maximum agency to the subject. However, it is ill-suited for privacy models that rely on hiding the subject's record in a group of records, such as *k*-anonymity and its extensions, because those methods necessitate the clustering of the contributions of several data subjects. If we obviate this difficulty and want to pursue local anonymisation, randomised response and *local* DP are natural approaches. Unfortunately, the current literature on both approaches focuses on obtaining statistics on the data from subjects, rather than multi-dimensional full sets of anonymised microdata that are valid for exploratory analysis.

- *Anonymised data are difficult to merge and explore.* Even if subjects decide to accept centralised anonymisation by the controllers, none of the main families of privacy models manages to satisfy all the desiderata of big data anonymisation identified [66]: (i) *protection* against disclosure no matter the amount of background information available to the intruder; (ii) *utility* of the anonymised microdata for exploratory analyses; (iii) *linkability* of records corresponding to the same or similar individuals across several anonymised datasets; (iv) *composability*, that is, preservation of privacy guarantees after repeated application of the model or linkage of anonymised datasets; and (v) *low computational cost*. Utility and linkability are needed to empower users/data analysts, protection and composability are desired by subjects, and low cost is desired by controllers. On the other hand, it is hard for controllers holding datasets to engage in joint exploratory analysis of their data without disclosing these to other controllers. Note that cryptographic secure multiparty computation (MPC) [67] is of little use here, because it is intended for specific calculations planned in advance rather than exploratory analyses. Furthermore, while MPC ensures input confidentiality, it gives exact outputs and cannot protect against disclosure by inference (e.g. if the mean and the variance of a sample are jointly computed and the variance is very small, it can be inferred that the values of all subjects lie close to the mean).

- *Ad-hoc SDC methods.* In the current state of the art, each privacy model is satisfied using a specific SDC method (or a few specific ones). For example, *k*-anonymity is reached via generalisation or microaggregation, and DP via noise addition. Life would be easier if a unified masking approach existed that, under proper parameterisation, could be used with a broad range of privacy models.

- *Stream data and real-time analytics.* The potential of streaming analytics for research, policy analytics and business is huge; it is potentially a game changer. However, the additional confidentiality issues raised by streamed data are significant. This again is a relatively uncharted area. Work done with mobility data indicates one possible direction [68, 69], but given the likely importance of streaming data, the need for focused research in this area is urgent.

- *Radical linkability.* In the context of a data environment in which data fragments about population units abound, sometimes referred to the *data soup* [70], traditional record linkage probably needs reformulating. A more flexible concept of *entity resolution* is coming to fore [71, 72, 73, 74] and the use of graphical and clustering-based approaches is being investigated [75, 76]. Some focused thinking about what risk means in this context is needed. It seems likely that there is an opportunity here as well as new risks, as these types of advanced linkage processes can be used as risk measurement instruments [77].

## 4.3 SDC and machine learning

The term "machine learning" has numerous definitions and has migrated in meaning since its original use [78]. A current consensus would probably form around a definition such as *a computational system which improves its performance on some task (involving the processing of data) by the (partially or fully automated) processing of feedback about that performance*. The boundary between machine learning and statistics is fuzzy but it is certainly true that many data analysts now use machine learning tools as well as more orthodox statistical ones.

There is a two-way relationship between SDC and machine learning:

- On the one hand, SDC must preserve sufficient utility in the masked data for these to be still useful for machine learning, as well as knowledge extraction and exploratory analyses. This is especially necessary if SDC is applied to big data. The requirements here are subtly different from those needed for inferential statistics – the utility required to train a model may differ from that required to apply that model.
- On the other hand, machine learning in a broad sense can be useful for SDC. At present this is an underexplored area. In the rest of this section, we discuss the specific potential of machine learning in synthetic data generation.

### 4.3.1 Data synthesis using machine learning

The classical goal of machine learning is to build models and make predictions based on discovered patterns in data. In principle, this maps well onto the goals of data synthesis where, paradigmatically, the synthetic data is one draw from a posterior distribution generated using the original data. Recently, there have been attempts to synthesise data using deep learning techniques such as generative networks across a range of data types for example medical records [79], sensor data [80] and computed tomography images [81]. The results are certainly interesting; however, although they have recognised "privacy" concerns, they are not at present focused on the SDC use case.

Within the SDC framework, a machine learning approach to synthetic data generation using genetic algorithms has been developed which shows much promise [45, 55, 79].

Traditional methods for synthetic data production aim to produce some model of the original dataset, and then uses that model to generate the data. This can be viewed as selecting from the (very large) space of possible datasets one that is consistent with the model. This might be contrasted with the application of SDC methods which could be viewed as an operator to move through the space (away from the original data). GA approaches generate synthetic data by similarly moving through the space. However, whereas with SDC the movement is a by-product of the operator and one needs a completely different method (risk and utility test) to assess if one has arrived at a good enough place, GA data synthesis is explicitly goal-directed.

The advantages of the GA approach are twofold: (i) it allows optimising utility and disclosure risk within a single system (rather than having to face the choice between favouring utility and controlling risk ex-post on the one side, or limiting risk ex-ante with a privacy model on the other side) and (ii) rather than the generation process being based on a single model, one can optimise across a (in principle unlimited) basket of required analytical properties.

The initial GA population (set of candidate datasets) is typically either randomly generated or drawn from univariate distributions. One interesting feature, though, is that if one instead uses copies of the original data for the initial population then the problem looks like a classic SDC problem (where the utility is initially optimised but the risk is too high). This underlines the point that the boundary between data synthesis and SDC may be more of form than substance [34].

### 4.4 Frameworks to unify SDC

SDC has been so far characterised by a great diversity of masking methods, disclosure risk assessment approaches and privacy models. This makes it hard to compare the various approaches being proposed. Quite recently, there have been unifying efforts. We mention next: i) the permutation model and the maximum-knowledge intruder; and ii) differential confidentiality for synthetic data generation.

### 4.4.1 Unifying SDC: The permutation model and the maximum-knowledge intruder

A permutation model of anonymisation has been proposed, which shows that any masking algorithm can be viewed as a permutation of the original data plus a small amount of noise [83]. Thus, any masking can be basically modelled as permutation, and checking whether the amount of masking is sufficient for protection reduces to checking whether permutation is sufficient. This can be useful to provide anonymisation transparency, to define new privacy models and to measure risk and utility in a more unified way.

Using the permutation model, a worst-case intruder model with the so-called *maximum-knowledge intruder* who knows the entire original dataset and the entire anonymised dataset has been introduced [83]. His objective is to discover the correct linkage between original and anonymised records, that is, to discover the permutation that basically turns the original dataset into the anonymised dataset. Such an intruder is stronger than any intruder in the literature: since he knows the original dataset, external background knowledge is as irrelevant to him as it is in $\varepsilon$-differential privacy. His motivation is purely malicious, e.g. to tarnish the data controller's reputation. An open research avenue is to investigate how successful such an intruder can be at finding the linkage and verifying its correctness depending on the information he is given on the anonymisation method.

### 4.4.2 Differential confidentiality

A concept of *differential confidentiality* in the context of synthetic data generation which draws on the attribute or inference disclosure metric CAP described in Section 4.1 has been recently developed [64]. The question posed is: does the presence of a particular population unit in the original dataset allow stronger (higher probability) inferences to an intruder with any given prior knowledge about that unit than if the unit was not in the original dataset? The method allows the measurement of information leakage in a synthetic dataset. The method could also be applied to disclosure-controlled data.

### 4.5 SDC and anti-discrimination

Data mining is gaining societal momentum due to the ever-increasing availability of large amounts of human data, easily collected by a variety of sensing technologies. There are unprecedented opportunities and risks: a deeper understanding of human behaviour and how our society works is darkened by a greater chance of privacy intrusion and unfair discrimination. The individuals whose data are published may suffer discrimination in decisions made on them if data mining models (e.g. classifiers) are trained on data that are biased against certain protected groups (ethnicity, gender, political preferences, etc.).

Fortunately, it has been shown that a synergy exists between confidentiality protection and anti-discrimination protection. A methodology has been outlined [84] to obtain datasets for publication that are: i) confidentiality-preserving, that is, anonymised; ii) unbiased regarding discrimination; and iii) as useful as possible for learning models and finding patterns. The approach to simultaneously achieve confidentiality preservation and discrimination prevention uses generalisation as an SDC method. It turns out that the impact on the quality of data is the same or only slightly higher than the impact of achieving just confidentiality preservation.

If the above combines confidentiality preservation and discrimination prevention for data publication, the same can be achieved with knowledge extraction [85]. Consider the case in which a set of patterns extracted from the personal data of a population of individual persons is released for a subsequent use in a decision-making process such as granting or denying credit. First, the set of patterns may reveal sensitive information about individual persons in the training population and, second, decision rules based on such patterns may lead to unfair discrimination, depending on what is represented in the training cases. This describes a set of pattern sanitisation methods, one for each discrimination measure used in the legal literature, to achieve a fair publishing of frequent patterns in combination with two possible data transformations: one based on *k*-anonymity and one based on differential privacy. The proposed pattern sanitisation methods based on *k*-anonymity yield both confidentiality and discrimination-protected patterns, while introducing reasonable (controlled) pattern distortion. Moreover, they obtain a better trade-off between protection and data quality than the sanitisation methods based on differential privacy [85].

### 4.6     Local and collaborative approaches to SDC

Local anonymisation is an alternative paradigm in which each subject anonymises her own data before handing them to the data controller, who therefore does not need to be trusted. Several SDC methods can be applied locally (including generalisation/recoding and noise addition). On the other hand, there are methods specifically designed for local anonymisation that, in addition to helping subjects hide their responses, allow the data controller to get an accurate estimation of the distribution of responses for groups of subjects (for example, randomised response [86] and FRAPP [87]). There are also a number of methods aimed at obtaining differentially private preselected statistics through local anonymisation, like RAPPOR [88] and local DP [89, 90].

For privacy models relying on hiding subjects in a group (such as *k*-anonymity and its extensions *l*-diversity, *t*-closeness and others), local anonymisation is not suitable: forming clusters requires some collaboration between subjects. Rational anonymisation collaboration is sustainable due to the game-theoretic notion of co-utility that characterises situations in which the best strategy to attain one's goal is to aid others in attaining theirs [91]. Privacy is a *public good* that is naturally co-utile, in the sense that: *the more private stay the people among whom I hide, the more private I stay*. Specifically, in microdata data anonymisation via clustering, the confidentiality of a subject in a cluster depends on the confidentiality of the rest of subjects in the cluster: none of the subjects is interested in making any of the other subjects re-identifiable. Co-utility leads to protocols that work smoothly without external enforcing mechanisms.

A naïve co-utile collaborative anonymisation protocol has been given to achieve k-anonymity [92]: in step 1, subjects share the values of their QIs (that are not confidential) so that any of them can compute and publish clusters of *k* subjects based on their QIs; in step 2, subjects use anonymous channels to send the values of their confidential attributes together with the number of the cluster where they ought to be placed. In this way, a *k*-anonymous dataset is collaboratively obtained. Further research is needed to deal with malicious behaviours of subjects aimed at framing a target subject in order to learn the latter's confidential data (for example, by fabricating *k-1* false subjects with QI values similar to the ones of the target subject).

## 4.7 Linkage-synthesis models

The combination of record linkage and synthesis is an intriguing methodological possibility that has yet to be explored beyond conjecture [93]. There are numerous potential avenues that might be researched within this framework (including a mechanism for multi-source linkage) but a key potential use case is a methodology for linkage which circumnavigates the need for using TTPs (using direct identifiers).

Suppose that Alice owns dataset *A* (containing set of variables *{WX}*) and Bob owns dataset *B* (containing set of variables *{YZ}*) which may, or may not, contain units from *A*. Both Alice and Bob have access to *C* (*{XY}*) – say a census microdata file.

*C* can be used to enable Alice to produce a partially synthetic dataset consisting of the original *A* with synthesised values *Y'*. *Y'* could then be passed to Bob who could link it against the real records in dataset *B*. *Y'* would then be passed back to Alice with a serial identifier (ID) attached denoting the real record in *B* that best matched each synthetic record in *Y'*. This would be iterated multiple times. Alice could then identify from the information that Bob has returned the ID in *B* that is most frequently hit for each record in *A* ("the optimum links set"), which would be the presumed link. The process could also then be reversed with Bob synthesising and sending a synthetic *X'* to Alice, and Alice sending back IDs linked to that *X'*.

If we imagine that Alice and Bob are actually the data scientists working for data holding organisations and a researcher wants to analyse the combination of *AB*, then the net result of the above process would be a set of linked IDs which could be sent to a researcher (or research data centre) who could then be sent the two datasets which could then be linked. The key point is that at no point in this process have the direct identifiers for the data subjects been shared with a third party (trusted or not).

There would be complex meta-issues to be resolved (e.g. contradictions in the *A->B* and *B->A* links and many-to-one links) but these issues also represent leverage on identifying the correct links. Thus, this represents a complex and interesting, but directly applicable, research problem.

Another extension of this approach is where the goal of the process is not to produce a set of links but rather to use the process as a mechanism for enhancing the dataset owned by each party with a synthetic version of the data owned by the other party (which would be better than any synthetic data they could have produced without the mechanism). One simple option to explore is to assume that, for a given record in *A*, multiple imputation from the optimum link set will produce better results than multiple imputation using *A* and *C* only.

### 4.8 Automated data environment mapping

The difficulty with the current approach to data environment analysis (described in Section 3) is the time it takes to identify collection instruments and manually enter the metadata into the repository. This makes the approach non-viable for any realistic data dissemination activity with the possible exception of the Census (the methodology actually arose from the 2011 UK Census). Currently, each collection instrument is added to the repository as a list of variables and categorisation mappings (using standardised variable / category names that match those in the relevant ontology). The ontology is a graph that describes the relationships between the categories of a variable.

In order to generate a large repository that will provide a reasonable representation of the data environment that would have wide usability (and perhaps even more challenging to update that repository in near real time), the process must be automated. This would require several computational processes which mix Artificial Intelligence (AI) and web science:

1. Information retrieval – crawling the web to identify data collection instruments.

2. Text mining – identifying the variables on the collection instruments and their categorisations.
3. Ontology generation – identifying the relationships between categories so that matching possibilities can be identified.

This is at a pre-theoretical stage at present; pilot work needs to be carried out in order to justify the significant R&D investment required for producing the new system (and then ongoing resources to maintain such a system).

Textual information from collection instruments will need to be mapped to variables. For each variable instance encountered, the categories need to be mapped to the ontology that describes the relationships between the (previously encountered) categories of the variable. New variables/categories need to be identified so that the repository/relevant ontology can be updated. This would build on initial work which outlined the basic principles of key variable mapping [94].

Mapping questions to variables is a non-trivial natural language processing problem. A question and a set of possible responses do not necessarily map to a single variable (e.g. a 'General Health' question listing diseases where the respondent can select zero or more diseases from which they suffer). Even if the mapping is 1 to 1 there might be an additional category that is implied by not selecting any of the possible responses ('Are you X?', with a single tick box).

Iteratively constructing an ontology is also non-trivial. Inconsistencies between a new categorisation and an existing ontology are easily identified from the graph structure. Resolving the inconsistencies might involve correcting a mapping (a collection instrument might have an implied category that has not been recognised) or correcting the ontology and any previously added variable instances (e.g. the first variable instance added, where the ontology would be created from the set of identified categories, but the categories might not have been exhaustive). Inevitably there will be some cases that need to be flagged for human intervention. In order to be feasible, the proposed system requires that the proportion of cases where human intervention is required be low. This is what the proposed work seeks to demonstrate.

Information will be retrieved for easily identified and parsed collection instruments (say limited to HTML with <form> tags). These will be used to build a corpus of forms (collection instruments). A set of variables will be identified, and classifiers used to generate (probabilistic) mappings of form data to the set of variables. Ontologies will be represented as graphs, with inconsistencies being identified using specific graph properties (these graphs are already developed). The (probabilistic) mappings for variables and categories and the degrees of inconsistency with existing ontologies will be used to guide decisions (update ontology and add variable instance or flag for human intervention).

One approach that could be investigated is to manually classify a corpus of forms in order to seed the system with correct mappings and ontologies. These can be used to learn the classifiers. The system would then be tested against forms that were not used for generating the classifiers. This approach might reduce the levels of human intervention subsequently required. The important thing is that such intervention should become rarer as the size of the repository increases.

## 5. Concluding remarks

In this article we have explored the future of SDC; in this we have endeavoured to focus on immediate challenges and practical solutions whilst allowing ourselves some freedom to speculate. Given the hazardous nature of any horizon scanning activity, and given the aforesaid rapid pace of change, we are bound to have missed future possibilities and outcomes. However, we are confident that the possibilities and challenges that we have focused upon are potentially important and this leads us to a set of recommendations.

### 5.1 Recommendations

We recommend that the community should:

1. Evolve statistical disclosure control so that it can deal with the challenges brought by big data, namely:
   a. Empower data subjects against the crowd of untrusted data controllers by developing local, collaborative and transparent SDC methods.
   b. Empower data users and controllers by developing SDC methods that offer disclosure protection, utility for exploratory analyses, selectable linkability between anonymised data sources, composability and low computational cost.
   c. Unify the vast array of SDC methods and privacy models, by devising SDC methods that can satisfy several privacy models and by defining utility and risk measures that are valid no matter the privacy models and SDC methods used.
   d. Investigate the particular challenges presented by new forms of data, most notably data streams.
2. Integrate data sanitisation for privacy protection with data sanitisation for other purposes, such as anti-discrimination [84]. Finding synergies between different types of sanitisation may incur less information loss than independently pursuing the various sanitisation goals on the same data.
3. Carry out a thorough investigation of the potential for machine learning (including deep learning and optimisation methods) for confidentiality protection. This might include how to use big data to unobtrusively generate official statistics. Two specific cases:
   a. Pilot the possibility of producing an AI system to automate key variable mapping.
   b. Carry out a programme of research to enable the Government Statistical Service (GSS) to fully realise the potential of synthetic data. This research should consider synthesis methods, disclosure risk measurement and utility optimisation and the potential of modern AI techniques including deep learning.
4. Investigate how SDC methods may be adapted to learn from the insights provided by differential privacy.
5. Carry out work to formalise output disclosure risk assessment.

As a closing remark, our overarching recommendation is that this type of horizon scanning activity should be carried out regularly, for identified trends to be constantly reviewed and for old decisions to be frequently revisited.